\documentclass[aps,prl,twocolumn,showpacs,amsmath,amssymb]{revtex4}
\usepackage{graphicx}% Include figure files
\usepackage{dcolumn}% Align table columns on decimal point
\usepackage{bm}% bold math
\begin{document}

\title{
Coherent cavitation in the liquid of light}
\author{ \'Angel Paredes,  David Feijoo and Humberto Michinel}
\affiliation{\'Area de \'Optica, Departamento de F\'\i sica Aplicada,
\\ 
Universidade de Vigo, As Lagoas s/n, Ourense, ES-32004 Spain.
}

\begin{abstract}
%----------------------------   ABSTRACT  ------------------------------------
We study the cubic (focusing)-quintic (defocusing) non-linear Schr\"odinger equation
in two transverse dimensions. We discuss a family of stationary travelling waves, including
rarefaction pulses and vortex-antivortex pairs, in a background of critical amplitude.
We show that these rarefaction pulses can be generated inside a top-flat soliton
when a smaller bright soliton collides with it. The fate of the evolution strongly depends on the
relative phase of the solitons. Among
several possibilities, we find
that the dark pulse can re-emerge as a bright soliton.

\end{abstract}
%-----------------------------------------------------------------------------

\pacs{42.65.Jx, 42.65.Tg, 05.45.Yv, 02.30.Jr}

\maketitle

%--------------------------------------------------------------------------------

\paragraph{Introduction.-}

%I.\emph {Introduction.-}
The cubic (focusing)-quintic (defocusing) non-linear Schr\"odinger equation (NLSE) reads:
\begin{equation}
i\partial_z \psi = - \nabla_T^2 \psi - (|\psi|^2 - |\psi|^4)\psi,
\label{eq1}
\end{equation}
where $ \nabla_T^2$ is the Laplacian in $d$  dimensions
and the absolute values of the
coefficients have been taken to unity without loss of generality.
Eq. (\ref{eq1})
is a paradigmatic model to describe fluids in which an attractive interaction becomes
repulsive at short distances. It has been applied to superfluids \cite{superf}, 
Bose-Einstein
condensates \cite{bec}, nuclear matter
 \cite{nuclear} or plasmas \cite{plasmas}, among other fields \cite{panova}. 
 In  non-linear optics \cite{nlo}, $d=2$ 
 when dispersion effects are negligible.
Light beams acquire peculiar properties and can reach a phase with qualitative and
quantitative similarities to a liquid, as theoretically discussed in
\cite{dripping}. 
Media with non-linear refractive indices resembling (\ref{eq1})
include chalcogenide materials \cite{chalco} among others \cite{caplan}.
Following the idea of \cite{coherent-media} of using engineered coherent media,
the experimental realization of this liquid of light has been achieved recently 
in a gas of sodium \cite{experiment}.

The liquid-like features are related to the existence of
stable solitons and vortices \cite{top-flat-vortices} in $d=2$.
For large   $\int |\psi|^2 dS$, these 
solitary
waves become {\it top-flat}: there is a  region where
$\psi \approx \exp(i\,\beta_{cr}\,z) \Psi_{cr}$ and $\psi \approx 0$ elsewhere.
This is reminiscent of instanton interpolation between vacua \footnote{We thank
Albert Ferrando for suggesting this analogy.}, where instanton action is identified
with surface tension. The critical amplitude and propagation constant are \cite{critical}:
$
\Psi_{cr}=\sqrt3/2$, $\beta_{cr}=3/16
$.

In this Letter, we analyse  travelling wave solutions  in
a background of critical amplitude. These bubbles generated and evolving within a fluid 
are a realization of the phenomenon of cavitation.
In the liquid described by Eq. (\ref{eq1}), cavitation 
can take place when
flowing past an obstacle \cite{cavitation} or by explosion \cite{explosion}.
Alternatively, we discuss a
situation in which the coherence properties of the NLSE allow interference, providing a
scenario with essential differences to regular liquids.

We find a family of travelling waves  of constant velocity
\footnote{We loosely use the term velocity even if  the coordinate
associated to evolution of the NLSE is propagation distance rather than time.} and shape, including
rarefaction pulses
 (moving bubbles without vorticity)
 and vortex-antivortex pairs. These computations  follow the
seminal work \cite{jr1} (see also \cite{bethuel}) for cubic NLSE and its  generalizations 
\cite{chiron1,chiron3}. Then, we simulate collisions of a soliton of small power and size
with a top-flat soliton. 
Different outcomes are possible, depending on their relative phase and speed:
both solitons may merge in a single droplet, the impinging
soliton may bounce back or a rarefaction pulse of the aforementioned family may be generated.
This dark excitation sometimes re-emerges as a bright solitary wave.

In view of these qualitatively distinct results, it is conceivable to use
the present set-up for interferometric purposes or optical switching. For instance, it may be useful to measure
relative phases and speeds in experiments of the liquid of light
as those described in \cite{experiment}, and to determine the coherence of the beams.
This procedure to generate rarefaction pulses could find interesting applications
such as studying their interaction with vortices \cite{berloff1}, their stability properties when
embedded in three dimensions \cite{berloff2} (with  second order dispersion)
or their dynamics in the presence of smooth inhomogeneities \cite{Smirnov}.

\paragraph{Travelling waves.-}

We look for stationary  solutions of (\ref{eq1}) moving at constant speed $U$
in  the $x$-direction. 
In order to be form-preserving,
$\psi$ takes the form $ \psi(x,y,z)=e^{i\,\beta_{cr}\,z}\Psi(x-U\,z,y) $ \cite{jr1}.
Defining a new coordinate $\eta=x-Uz$, the partial differential equation (PDE) (\ref{eq1}) for 
$\psi(x,y,z)$ is reduced to a PDE for $\Psi(\eta,y)$ valid for any $z$. For instance we can take
$z=0$ where $\eta=x$ and write:
\begin{equation}
 iU\partial_x \Psi= \nabla^2\Psi + \left(|\Psi|^2 - |\Psi|^4-3/16\right)\Psi
 \label{eq2}
 \end{equation}
where $\nabla^2=\partial^2_x+\partial_y^2$.
Boundary conditions are
$
\Psi \to \Psi_{cr}=\sqrt3/2$, as $r^2 =x^2 + y^2 \to \infty
$
in any direction.
Momentum and energy are defined by:
\begin{eqnarray}
p &=& \frac{1}{2i}\int \left[(\Psi^* - \Psi_{cr})\partial_x \Psi
-(\Psi - \Psi_{cr})\partial_x \Psi^* \right] dxdy
\nonumber
\\
E&=& \int |{\bf \nabla} \Psi|^2 dxdy +
\frac13
\int |\Psi|^2\left(|\Psi|^2-\Psi_{cr}^2   \right)^2 dxdy
\label{pE2}
\end{eqnarray}
If the
derivative is taken over the family of solutions, $U=\partial E/\partial p$ holds. 
$\Psi$ is even under $y \to -y$ whereas its imaginary (real)
part is odd (even) under $x \to -x$.

All solutions are subsonic $U < U_0$ where
$U_0=\sqrt3/2$ is the long-wavelength speed of sound. In the
transonic limit $U\to U_0^-$, there is an explicit solution 
\cite{jr1,chiron2}. 
We split real and imaginary parts as $\Psi = f+i g$ and introduce
stretched coordinates
$
\xi = \epsilon^2 y\,,\ \eta= \epsilon\,x
$. Then:
%there is a solution of the form
\begin{equation}
f= \Psi_{cr}+\epsilon^2 f_1 + \epsilon^4 f_2 + \dots ,
\quad g=\epsilon g_1 + \epsilon^3 g_2 + \dots
\end{equation}
with $U= U_0+\epsilon^2 U_1 + \epsilon^4 U_2 + \dots\nonumber$.
Expanding in $\epsilon$, one finds
$
f_1 = -\frac{1}{\sqrt3}g_1^2 + \frac{2}{\sqrt3} \partial_\eta g_1
$ and a Kadomtsev-Petviashvili equation for $g_1$:
\begin{equation}
\frac{4U_1}{\sqrt3}\frac{\partial^2g_1}{\partial \eta^2}-
\frac{\partial^2g_1}{\partial \xi^2}+
\frac43 \frac{\partial^4g_1}{\partial \eta^4}-16 
\frac{\partial g_1}{\partial \eta}
\frac{\partial^2g_1}{\partial \eta^2}=0
\end{equation}
which, for $U_1 = -\sqrt3/4$, is solved by:
\begin{equation}
g_1 = \frac{-2\eta}{\eta^2 + \xi^2 +4}\,,\qquad
f_1= - \frac{4(4+\xi^2)}{\sqrt3(\eta^2 + \xi^2 +4)^2}
\end{equation}
The value of $U_1$ can be
 rescaled with $\epsilon$, but its sign is important since it shows that
 $U<U_0$
in this limit. This provides the low momentum limit of the family of travelling
waves $E\approx \pi\,\epsilon$, $p=2\pi\,\epsilon/\sqrt3$, which corresponds to rarefaction
pulses. Solutions for $p\to \infty$,
 correspond to a well-separated vortex-antivortex pair with $U\to 0$.

\begin{figure}[htb]
\centerline{\includegraphics[width=.5\textwidth]{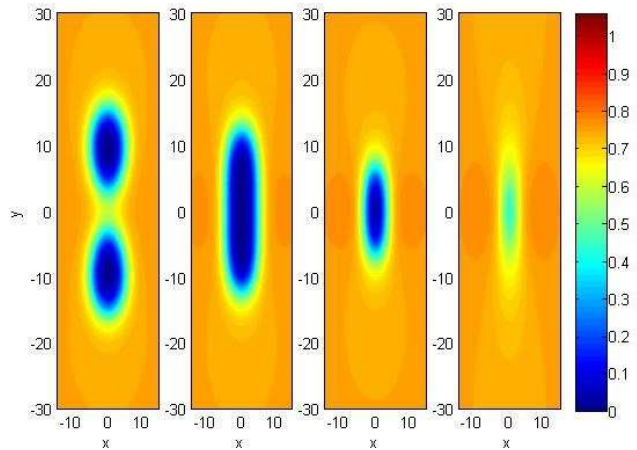}}
\caption{ (Color online) Contourmaps of $|\Psi|^2$ for four
travelling waves. From left to right: ($U\approx 0.11$, $p\approx 80$, 
$E\approx 20.7$);
($U\approx 0.21$, $p\approx 35.3$, $E\approx 12.5$);
  ($U\approx 0.35$, $p\approx 10.9$, $E\approx 6.1$);
  ($U\approx 0.71$, $p\approx 2.6$, $E\approx 2.1$). The color scale also 
  applies to Figs. \ref{fig2}-\ref{fig7}.
}
\label{fig1}
\end{figure}

We have computed numerical approximations to this family of solutions, following 
\cite{chiron3}: define a functional depending on a parameter $\mu=p+U$, such
that solutions of (\ref{eq2}) are its minimisers. Given a judicious initial ansatz for $\Psi$, they are computed by
a heat flow in some auxiliary time coordinate, see \cite{chiron3} for details. We found a first solution for large $\mu$
 by using a vortex-antivortex ansatz in the spirit of \cite{berloffvortex} and then used it
as starting point to go over the family by a subsequent iteration $\mu \to \mu - \delta \mu$.
We have cross-checked the numerical approximations
got at the end of each heat flow
 by requiring that the following virial identities, analogous to
\cite{jr1}, \cite{jr2}, are satisfied
within a $3\%$ in the finite difference scheme:
\begin{eqnarray}
E&=&p\,U+\int\left(|\Psi|^2 - \Psi_{cr}^2 
\right)\left( \frac{\sqrt3}{4}  \Big(|\Psi|^2 - \frac14 \Big)
(\Psi + \Psi^*) \right. \nonumber \\
 &+& \left.
-\frac23 |\Psi|^4 
\right)
dxdy =2 \int |\partial_x \Psi|^2 dxdy \nonumber
\\
p\,U&=&\frac23 \int |\Psi|^2\left(|\Psi|^2 - \Psi_{cr}^2 
\right)^2 dxdy
\label{ident}
\end{eqnarray}
Four examples are displayed in 
Fig. \ref{fig1}.
The dispersion relation is presented in Fig. \ref{fig6} below.
Arguments in \cite{chiron1} prove that these solutions are orbitally stable.
The stability of the family in \cite{jr1} was first analysed in
\cite{berloff_roberts}.

Static ($U=0$) bubble solutions in cubic-quintic media were found
long ago \cite{bubbles}, even if they are unstable \cite{unstable-bubbles}. 
The travelling waves presented here are not finite velocity generalizations of these bubbles
(such a generalization in one-dimension was discussed in \cite{moving-bubbles}).
 We have checked that the
static bubbles only exist for $\beta > \beta_{cr}$ and therefore the boundary conditions
used in this section do not match those of any quiescent bubble.

\paragraph{Rarefaction pulses from soliton-soliton collisions.-} 
We now analyse  the collision of two solitons of markedly 
different sizes by numerically
solving Eq. (\ref{eq1}). For these simulations, we use a standard split-step pseudo-spectral Fourier method ---
beam propagation method, see {\it e.g} \cite{poon} --- in a lattice of $512 \times 256$ points.
Even if it is a first order algorithm in  $\Delta z$, we have used
a fourth order Runge-Kutta to compute the evolution associated to the non-linear terms to avoid problems
with high non-linearities. Being an
explicit method, it is only conditionally stable and a sufficiently small step $\Delta z$ must be taken,
see \cite{instabBPM} for more details on this issue.
In order to write down the initial conditions, we notice that there exist smooth bright soliton
solutions of the form $e^{i\,\beta_s z}\,\Psi_s(r)$ with $\beta_s < \beta_{cr}$ and 
$\lim_{r\to \infty} \Psi_s(r) = 0$. Taking two of these and defining two radial coordinates
with respect to each initial position, $r_j^2=(x-x_{j,0})^2+y^2$ for $j=1,2$:
\begin{equation}
\psi|_{z=0}=\Psi_{s,1}(r_1) + \Psi_{s,2}(r_2)\exp\left(i\,\frac{v\,x}{2} +i\, \phi_i\right)
\end{equation}
corresponds to two separate solitons, the second one 
with initial velocity $v$. An initial relative phase $\phi_i$ has been included.
Fixing the initial solitons and their positions, we can study the dynamics as a function of 
$v$ and $\phi_i$. As an illustrative example, in all the following plots
 we take $x_{1,0}=-22$, $x_{2,0}=130$  and solitons 1 and 2 to
be those with $\beta_{s,1}\approx 0.1856$,  $\beta_{s,2}\approx 0.15$, corresponding to total power
$\int |\Psi_{s,1}|^2 dS \approx 3.1\times 10^4$, $\int |\Psi_{s,2}|^2 dS \approx 86$ and radii $R_{s,1} \approx 115$,
$R_{s,2} \approx 6.7$.

\begin{figure}[!htb]
\includegraphics[width=0.23\textwidth]{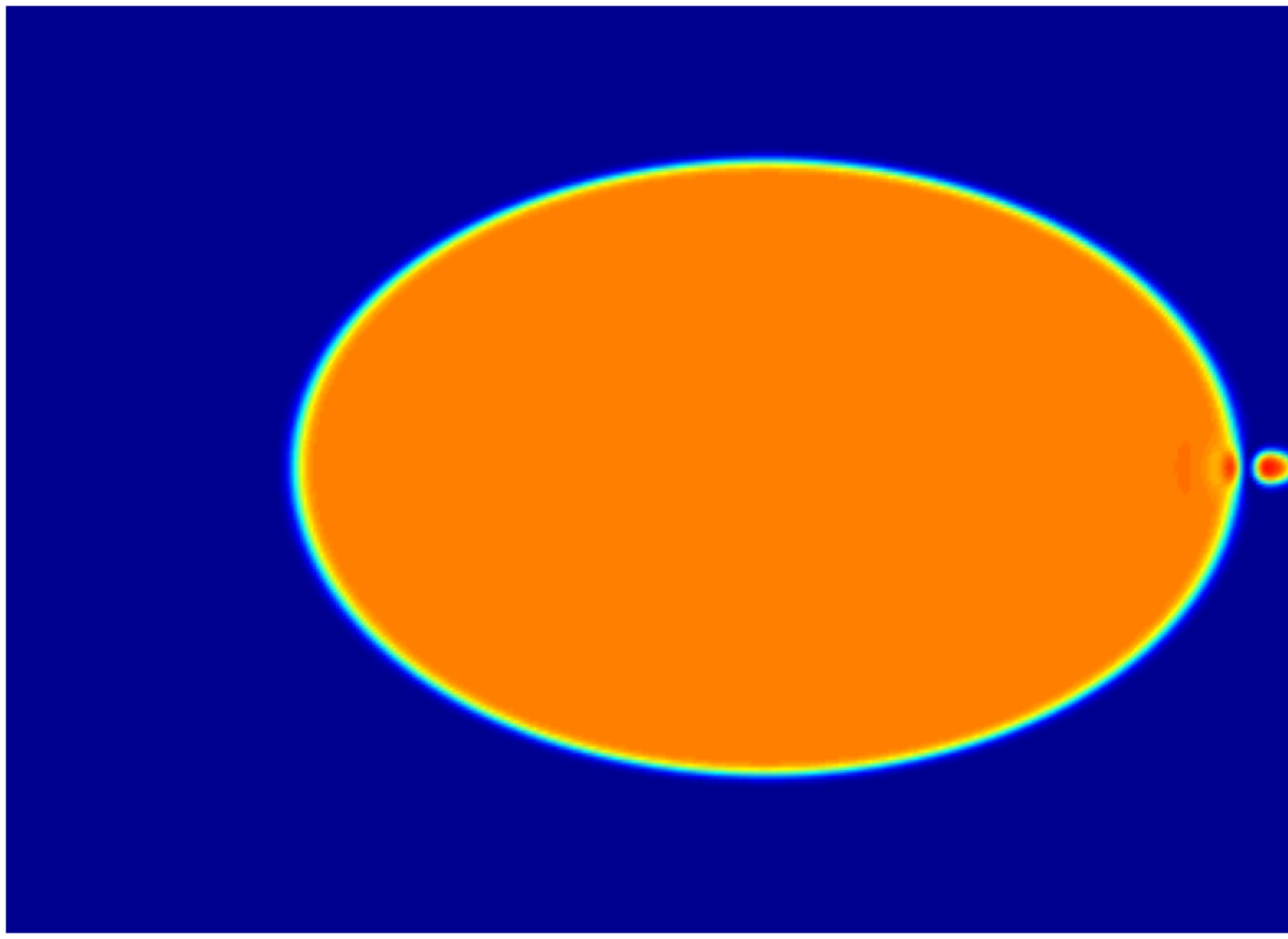}
\includegraphics[width=0.23\textwidth]{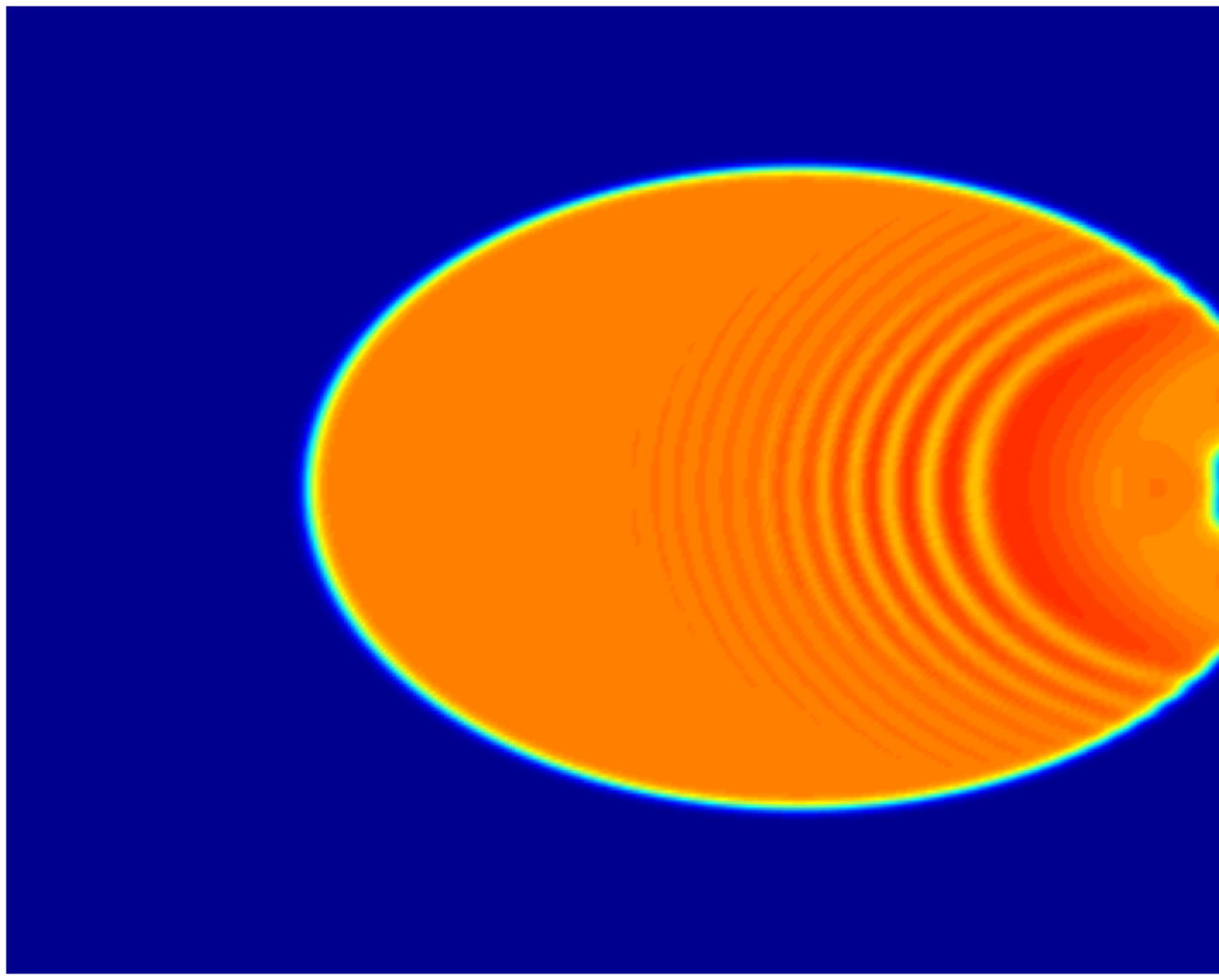}\\
\includegraphics[width=0.23\textwidth]{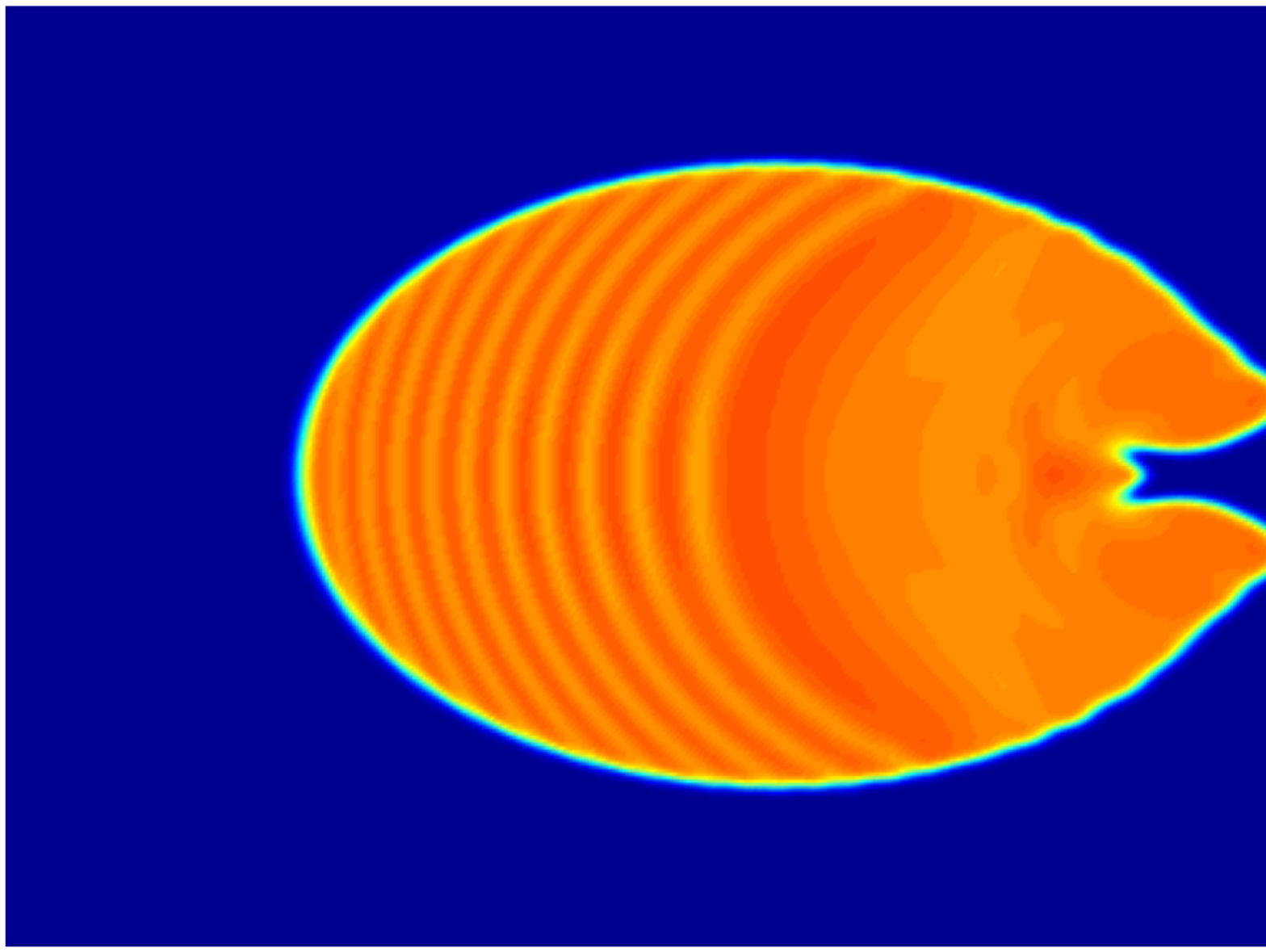}
\includegraphics[width=0.23\textwidth]{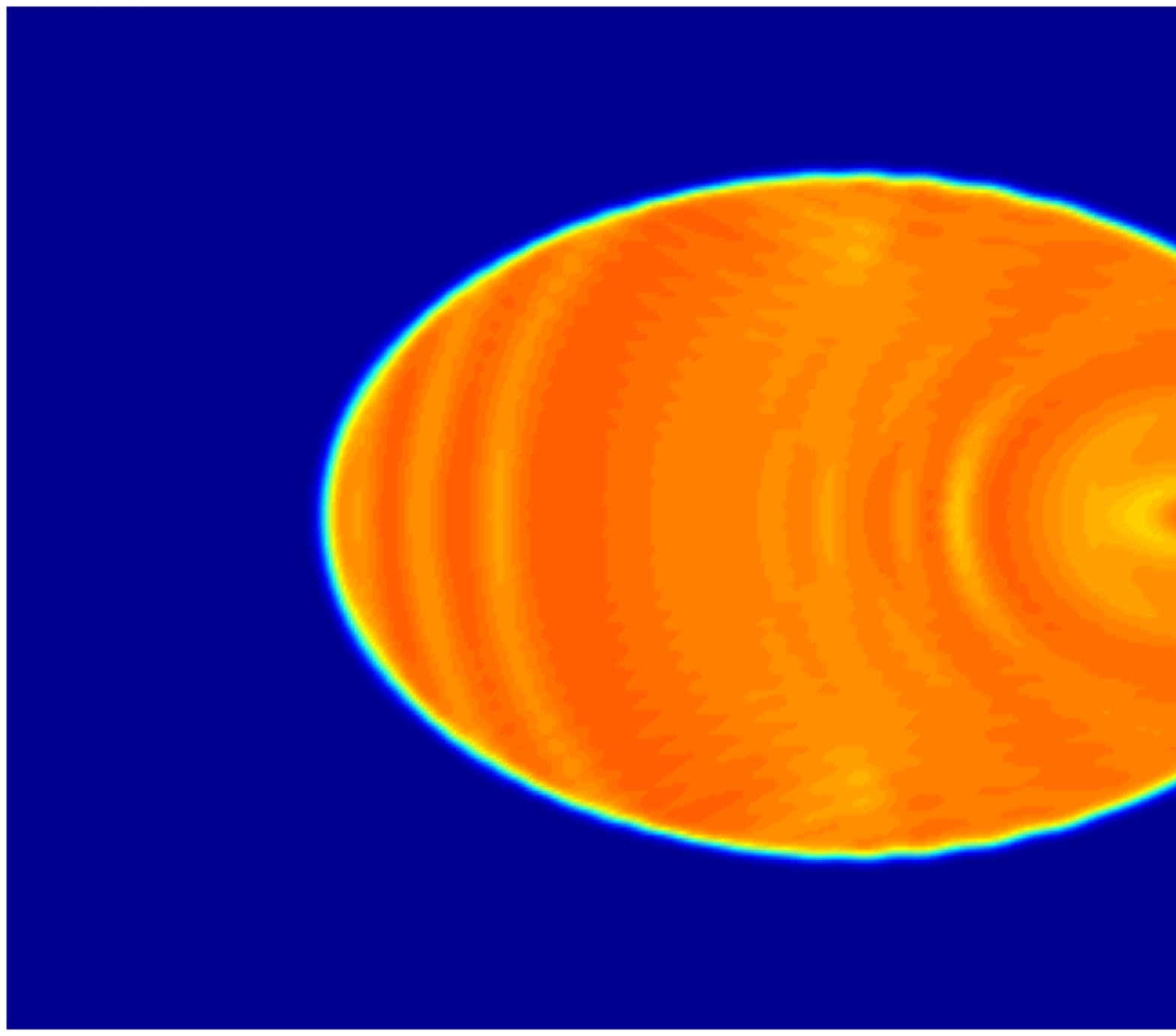}
\caption{ (Color online) Evolution of $|\psi|^2$ with initial conditions
$v=-1$, $\phi_i=\pi$. The four plots correspond to $z=30$, $z=90$, $z=150$, $z=210$, respectively.
Here and in Fig. \ref{fig5}, the horizontal axis is the $x$-direction,
$x \in (-205,205)$ whereas the vertical axis is $y\in (-172,172)$.
}
\label{fig2}
\end{figure}

\begin{figure}[!htb]
\includegraphics[width=0.23\textwidth]{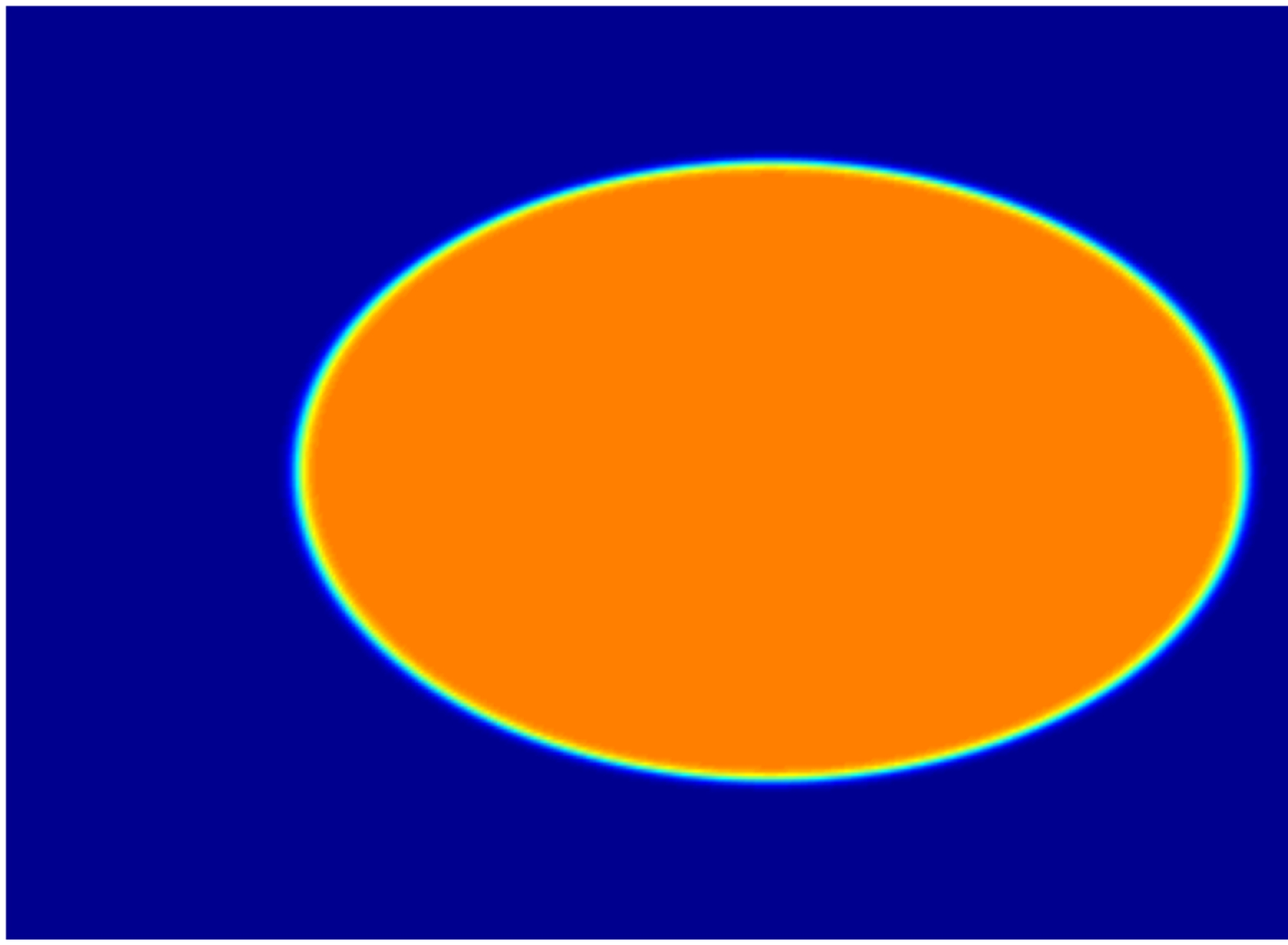}
\includegraphics[width=0.23\textwidth]{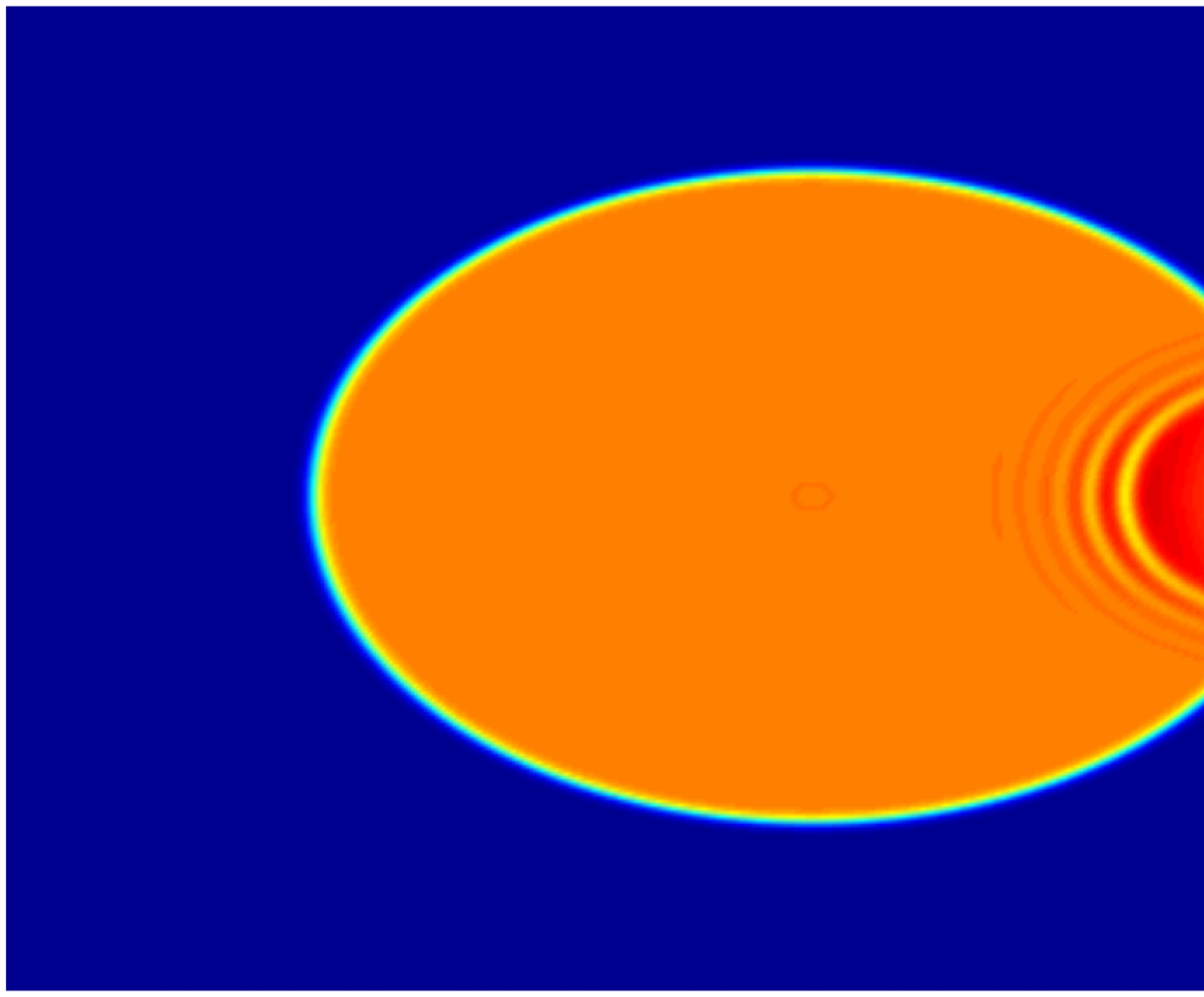}\\
\includegraphics[width=0.23\textwidth]{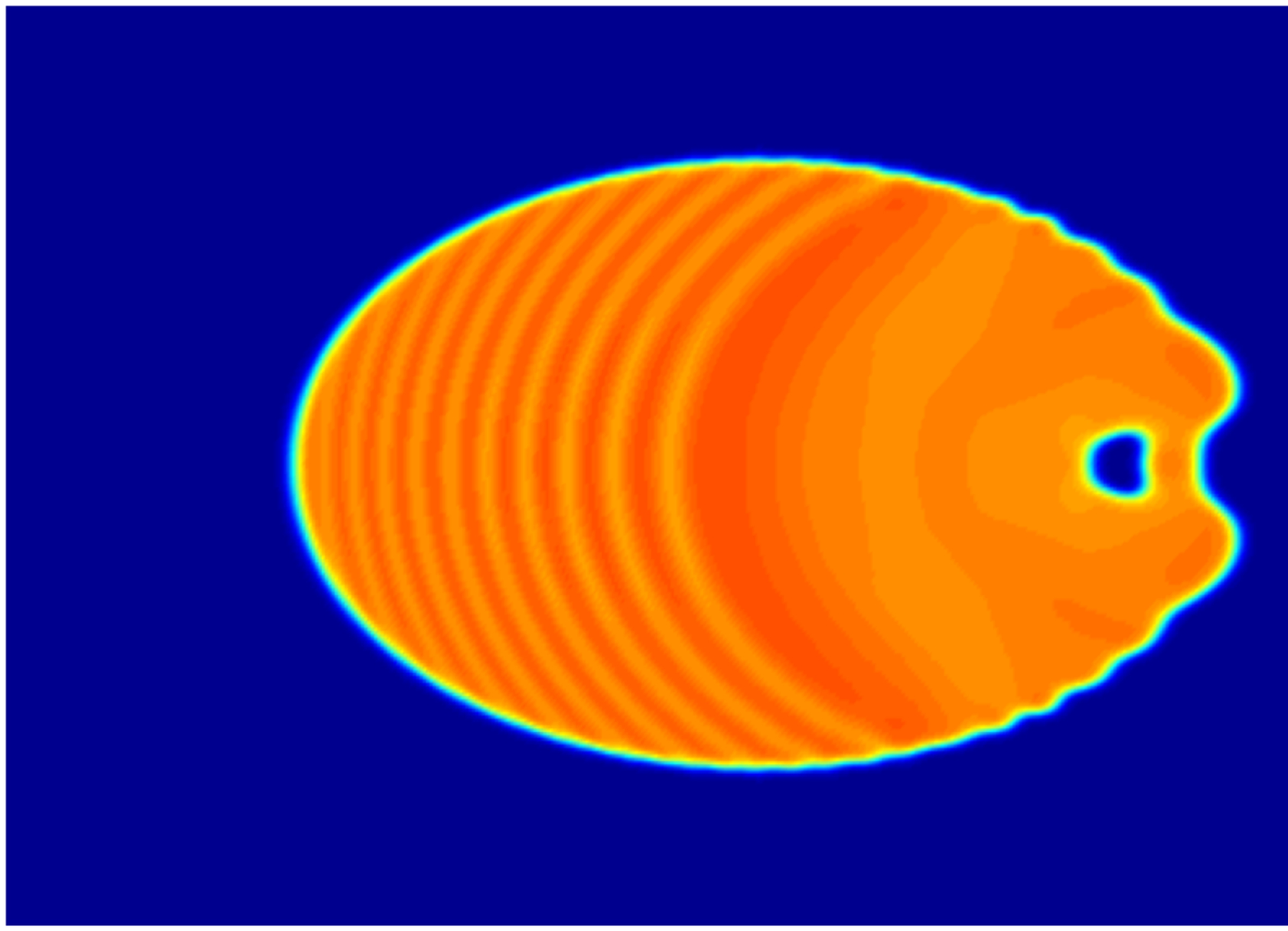}
\includegraphics[width=0.23\textwidth]{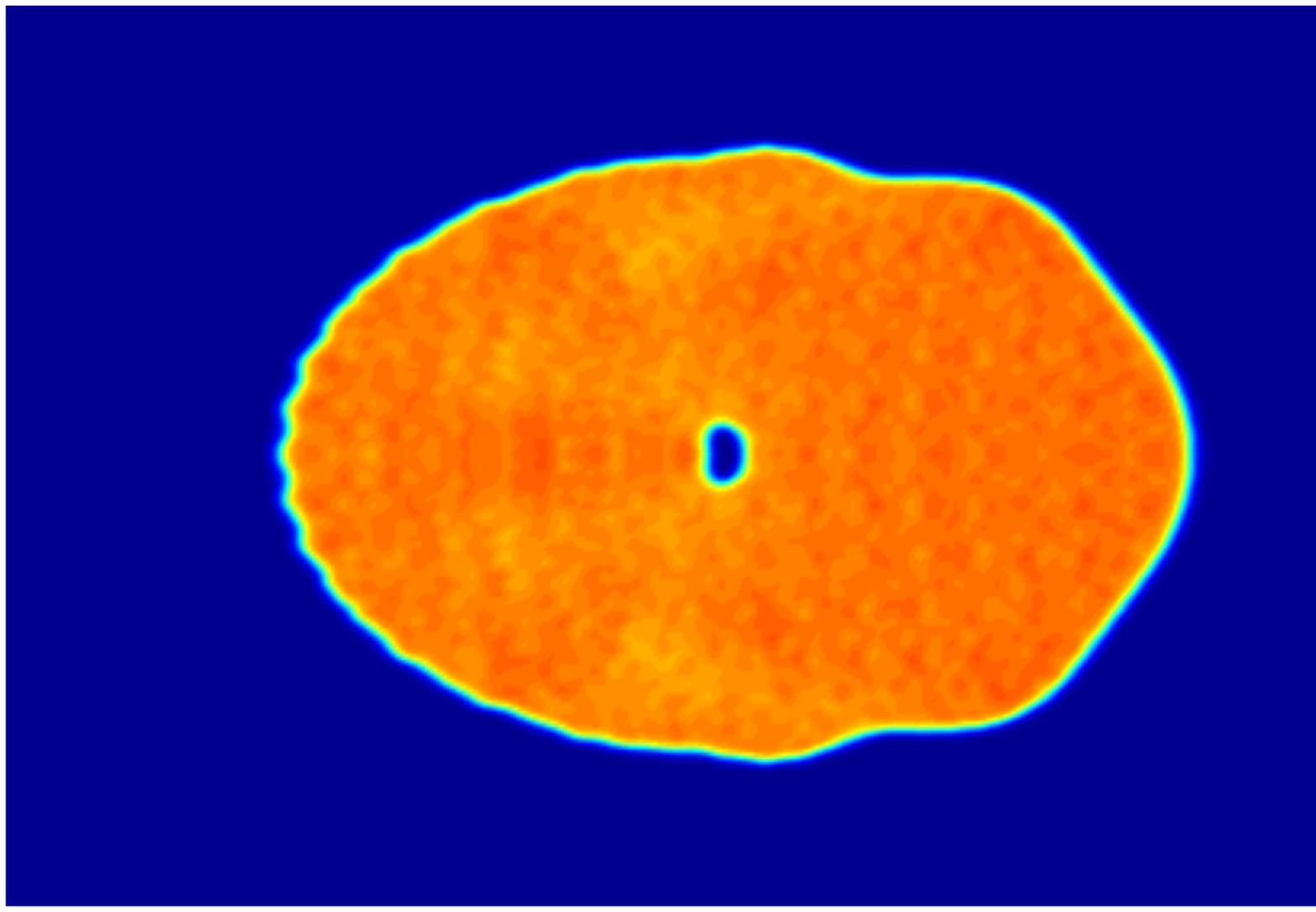}\\
\includegraphics[width=0.23\textwidth]{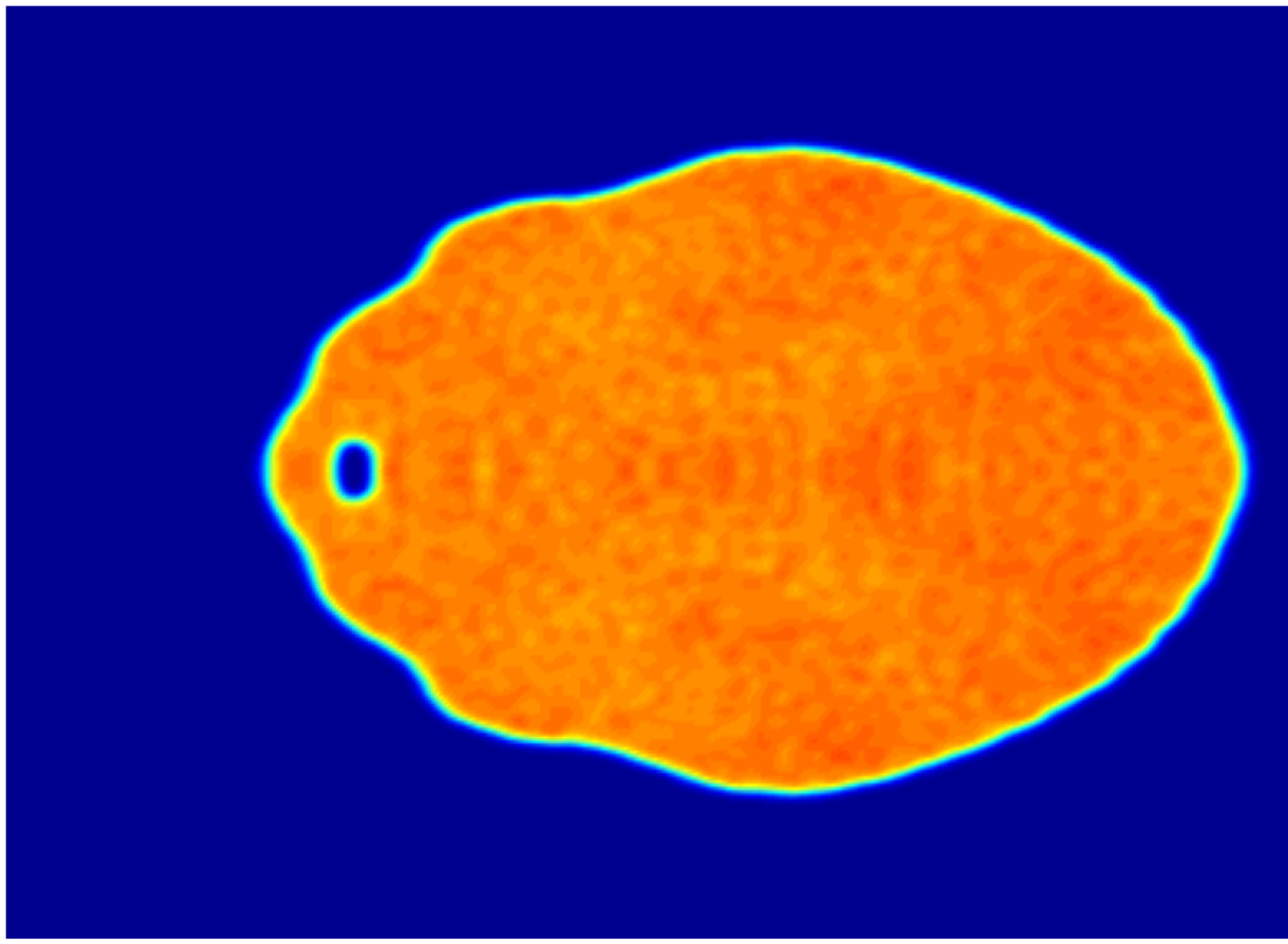}
\includegraphics[width=0.23\textwidth]{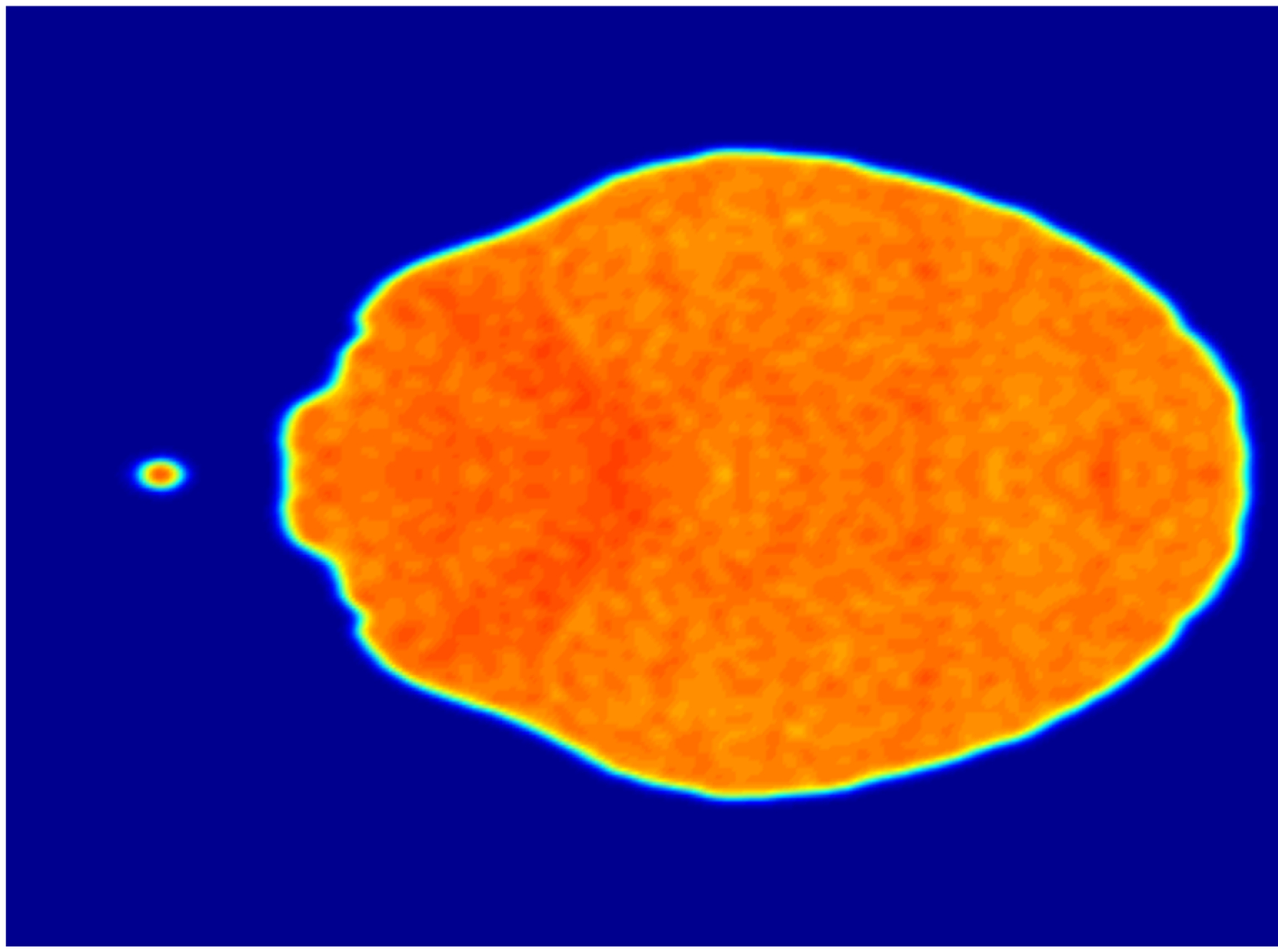}
\caption{ (Color online) Evolution of $|\psi|^2$ with initial conditions
$v=-1$, $\phi_i=0.7\pi$. The six plots correspond to $z=15$, $z=60$, $z=150$, $z=525$,
$z=900$, $z=1050$, respectively.
}
\label{fig5}
\end{figure}

%Figures \ref{fig2} and \ref{fig3} depict two qualitatively different outcomes.
In Fig. \ref{fig2}, both droplets coalesce into one, and the impact results in the excitation
of surface and body waves. 
In Fig. \ref{fig5}, a 
void is generated inside the droplet,
which subsequently moves with constant velocity
and eventually exits the top-flat soliton re-transformed in a bright soliton.
This bright-dark-bright excitation conversion, reminiscent of the one-dimensional case
\cite{1dexchange}, opens interesting possibilities. 
For fixed power of the initial
solitons, the re-emergence of the small bright
soliton only happens for certain ranges of initial soliton velocity and relative phase. Thus,
it may be a useful probe to  determine these parameters and/or the validity of the cubic-quintic 
model in a particular situation.

\begin{figure}[htb]
\includegraphics[width=0.45\textwidth]{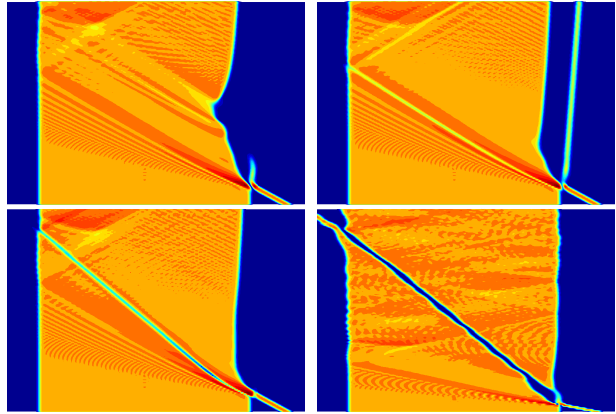}
\caption{ (Color online) 
We depict the evolution in $z$ of $|\psi|^2(x,y=0)$ of the simulations of
Figs. \ref{fig2} (top left) and \ref{fig5}
(bottom right). 
The top right figure is for initial conditions
$v=-1$, $\phi_i=1.3\pi$ and the bottom left for $v=-1$, $\phi_i=1.6\pi$
The horizontal axis corresponds to
$x \in (-205,205)$. The vertical axis represents evolution from $z=0$ to
$z=400$, except for the top right figure in which $z \in (0,1100)$.
}
\label{fig7}
\end{figure}

In Fig. \ref{fig7}, the evolution of $|\psi|^2(x,y=0)$
is plotted for the cases depicted above and for two other cases with
different qualitative outcomes. On the bottom left, the small soliton bounces back
--- also in Fig. \ref{fig2} part of 
the energy is reflected, although it is insufficient to form a bright soliton.
On the bottom right, a fainter rarefaction pulse is formed but dies away when reaching 
the border of the top-flat soliton.
Indeed, some rarefaction pulses propagate nearly undistorted
for a distance but eventually fade away.
This happens because the excitation is not exactly the stationary solution or because the medium
is not infinite and constant. The pulse can vanish when reaching the border of the top-flat
soliton or due to interaction with sound waves which also stem from the
original collision.
It can be appreciated that
 rarefaction 
  pulses move with constant velocity $U$, always slower than sound waves
  (in Fig. \ref{fig7}, velocity is 
the angle with respect to the vertical
axis).
  The relation between $U$ and $v$ is non-trivial:
depending on the relative phase, the larger is the $E$ of
the bubble, the smaller is its $U$.

Initial conditions of the aforementioned processes
 only differ in $\phi_i$. It is apparent that  coherence plays an essential role in the phenomena discussed. 
Apart from perturbations qualitatively similar to those happening when droplets of regular liquids collide,
there can be destructive interference near the surface of the large droplet.
The non-linear interactions allow this disturbance to evolve into a stationary bubble, resembling a vapour cavity
inside a liquid. In view of this analogy, we call this process {\it coherent cavitation}.
To further elaborate on this point, we write the relative phase before the solitons meet:
\begin{equation}
\phi_{rel} = -\phi_i - (\beta_{s,2} - \beta_{s,1} - \frac{v^2}{4})z - \frac{v\,x}{2}
\label{phase}
\end{equation}
Rarefaction pulses only appear above a limiting velocity
$|v| > v_{lim}$.
For the initial
solitons of Figs. \ref{fig2}-\ref{fig7}, $v_{lim}\approx 0.22$. The value of
$v_{lim}$ is larger for smaller incoming solitons.
For $|v|<v_{lim}$, the bounce is the most probable outcome.
 For $|v|>v_{lim}$, it is the formation
of a rarefaction pulse, which therefore appears under fairly general initial conditions. 
The cases in which it is not generated 
can be understood, at least qualitatively, by considering Eq. (\ref{phase}). A rarefaction
pulse appears unless $\phi_{rel}$ is around an integer multiple of $2\pi$ at the 
collision. As an example, for the case of the figures,
we have checked that the mentioned behaviour is observed in a considerable range of velocities
$0.25 \lesssim |v| \lesssim 3.5$ by 
inserting in Eq. (\ref{phase})
 $z=30.4/|v|$ (where 30.4 is the initial soliton-soliton distance) and
$x=90.6$, which is the position near the soliton edge where typically the larger $|\psi|^2$ disturbance
is created after the collision (thus $\phi_{rel}\approx 1.08/|v|+52.9|v|-\phi_i$).

In order to check that these bubbles in motion 
should be identified
with the family of travelling waves found above (Fig. \ref{fig1}), we have 
performed a series of numerical experiments and
compared their dispersion relation in Fig. \ref{fig6}.
Eqs. (\ref{pE2})
with two modifications were used to compute $p$, $E$. First, since the numerical integration gives
$\psi$ rather than $\Psi$, we have introduced $\Psi=e^{-i\beta_{s,1} z} \psi$.
Second,  the integration range cannot be taken to infinity. Our convention
has been to perform the integral in a rectangle in the $x-y$ plane defined as follows:
look for the positions $x_l$, $x_r$ where $|\psi|$ drops to $\Psi_{cr}/2$ at $y=0$.
The integration limits are $x_l - \Delta x < x <  x_r+\Delta x$,
$|y| < 0.82(x_r - x_l)$,
 where we 
have fixed $\Delta x  \approx 4$. This computation
can be done for different values of $z$ for the same
pulse. Results do not 
significantly differ as long as the rarefaction pulse is well within the large droplet
--- at least in a certain range of $z$.
As  expected \cite{review}, this dark solitonic excitation retains its identity
even if the background has finite extent and is not perfectly stationary.

\begin{figure}[htb]
{\centering \resizebox*{0.9\columnwidth}{0.545\columnwidth}{\includegraphics{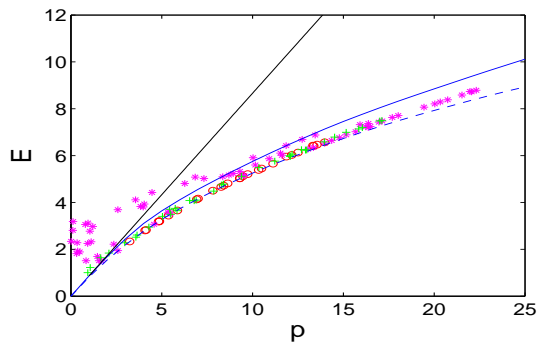}}}
\caption{ (Color online) $E$ vs. $p$ for several rarefaction pulses generated with different values
of $v$ and $\phi_i$ using  three pairs of 
initial solitons --- magenta asterisks for the case of Figs. \ref{fig2}-\ref{fig7}, green crosses 
for $\int |\Psi_{s,1}|^2 dS \approx 2.8\times 10^4$,
 $\int |\Psi_{s,2}|^2 dS \approx 42$ and 
red circles for $\int |\Psi_{s,1}|^2 dS \approx 
3.1\times 10^4$, $\int |\Psi_{s,2}|^2 dS \approx 33$.
 The solid and dashed curves correspond to the family of 
travelling waves of Fig. \ref{fig1} --- for the dashed one 
 the domain in the integrals  (\ref{pE2}) 
is cut 
 as described above. The black solid line
corresponds to sound speed $E=U_0 p$.
}
\label{fig6}
\end{figure}

Larger impinging solitons can generate larger bubbles (greater $p$, $E$). They also disturb more the top flat
soliton, explaining why some magenta asterisks deviate from the stationary dispersion 
relation for small $p$.

\paragraph{Conclusion.-} 
We have found  noteworthy qualitative behaviours in the  cubic-quintic
NLSE, including rarefaction pulses generated by soliton-soliton interference. 
We have shown that these bubbles can be identified with a family of stationary solutions. 
In view of the numerous applications of the model (\ref{eq1}) 
and the interest raised by the recent experiments \cite{experiment}, we hope that
our results might inspire novel experiments of non-linear optics in atomic gases and, possibly,
in other fields.

%\subsection*{{ Acknowledgements}}

%-------------------------------------------------------------------------
\paragraph{Acknowledgements.-}
We thank A. Ferrando, J. R. Salgueiro and D. Tommasini for discussions.
The work of A. Paredes is supported by the Ram\'on y Cajal programme. 
The work of D. Feijoo is supported by the FPU Ph.D. programme.
The work of A. Paredes and D. Feijoo is also supported by Xunta de Galicia through grant EM2013/002.

%--------------------------------------------------------------------------

\end{document}